\documentclass[sigconf]{acmart}

\newcommand{\sysname}{APINT }
\usepackage{algorithm}
\usepackage{algpseudocode}
\usepackage{amsmath}
\usepackage{indentfirst}
\usepackage{graphicx}
\usepackage{caption}
\usepackage{tikz}
\usepackage{enumitem}
\usepackage{titlesec}

\newcommand*\circled[1]{\tikz[baseline=(char.base)]{
            \node[shape=circle,draw,inner sep=1pt] (char) {\scriptsize #1};}}

\AtBeginDocument{%
  }

\copyrightyear{2024}
\acmYear{2024}
\setcopyright{rightsretained}
\acmConference[ICCAD '24]{IEEE/ACM International Conference on
Computer-Aided Design}{October 27--31, 2024}{New York, NY, USA}
\acmBooktitle{IEEE/ACM International Conference on Computer-Aided Design
(ICCAD '24), October 27--31, 2024, New York, NY, USA}
\acmDOI{10.1145/3676536.3676786}
\acmISBN{979-8-4007-1077-3/24/10}

\acmSubmissionID{1237}

\begin{document}

\title{APINT: A Full-Stack Framework for Acceleration of Privacy-Preserving Inference of Transformers based on Garbled Circuits}


      


\settopmatter{authorsperrow=4}
\author{Hyunjun Cho}
\email{h.cho@kaist.ac.kr}
\orcid{0009-0008-2378-852X}
\affiliation{%
  \institution{KAIST}
  \city{Daejeon}
  \country{South Korea}
}

\author{Jaeho Jeon}
\email{math15738@kaist.ac.kr}
\orcid{0009-0008-2250-6068}
\affiliation{%
  \institution{KAIST}
  \city{Daejeon}
  \country{South Korea}
}

\author{Jaehoon Heo}
\email{kd01050@kaist.ac.kr}
\orcid{0000-0003-1742-4275}
\affiliation{%
  \institution{KAIST}
  \city{Daejeon}
  \country{South Korea}
}

\author{Joo-Young Kim}
\email{jooyoung1203@kaist.ac.kr}
\orcid{0000-0003-1099-1496}
\affiliation{%
  \institution{KAIST}
  \city{Daejeon}
  \country{South Korea}
}

\begin{abstract}
As the importance of Privacy-Preserving Inference of Transformers (PiT) increases, a hybrid protocol that integrates Garbled Circuits (GC) and Homomorphic Encryption (HE) is emerging for its implementation. While this protocol is preferred for its ability to maintain accuracy, it has a severe drawback of excessive latency. 
To address this, existing protocols primarily focused on reducing HE latency, thus making GC the new latency bottleneck. Furthermore, previous studies only focused on individual computing layers, such as protocol or hardware accelerator, lacking a comprehensive solution at the system level.



This paper presents APINT, a full-stack framework designed to reduce PiT's overall latency by addressing the latency problem of GC through both software and hardware solutions. APINT features a novel protocol that reallocates possible GC workloads to alternative methods (i.e., HE or standard matrix operation), substantially decreasing the GC workload. It also suggests GC-friendly circuit generation that reduces the number of AND gates at the most, which is the expensive operator in GC. Furthermore, APINT proposes an innovative netlist scheduling that combines coarse-grained operation mapping and fine-grained scheduling for maximal data reuse and minimal dependency. Finally, APINT's hardware accelerator, combined with its compiler speculation, effectively resolves the memory stall issue. Putting it all together, APINT achieves a remarkable end-to-end reduction in latency, outperforming the existing protocol on CPU platform by 12.2$\times$ online and 2.2$\times$ offline. Meanwhile, the APINT accelerator not only reduces its latency by 3.3$\times$ but also saves energy consumption by 4.6$\times$ while operating PiT compared to the state-of-the-art GC accelerator.

\end{abstract}

\maketitle

\section{Introduction}

Transformer models have significantly advanced the fields of natural language processing (NLP), setting new performance benchmarks and becoming dominant in the area. This advancement has fueled the popularity of cloud services offering pre-trained Transformer models for NLP tasks, enabling clients to use these models without needing to own them. However, this convenience raises significant privacy concerns as it potentially exposes sensitive client data, like financial and health information, to cloud providers. Therefore, there is an urgent need for a new method that can prevent sensitive client data from being exposed to the server.

To address this issue, post-quantum cryptographic protocols such as Homomorphic Encryption (HE)~\cite{gentry2009fully, brakerski2014leveled} and Garbled Circuits (GC)~\cite{bellare2013efficient} are widely adopted for secure computation. HE enables calculations on encrypted data without decryption, preserving data confidentiality, but is limited to additions and multiplications, which hinders performance with nonlinear functions. In contrast, GC supports any operation expressible through the circuit composed of 2-input gates, allowing for more versatile applications despite its higher computational and memory demands.

Recent studies~\cite{juvekar2018gazelle, mishra2020delphi, garimella2023characterizing, zheng2023primer} have proposed a hybrid approach that combines the strengths of both protocols: using HE for linear functions and GC for nonlinear functions. This approach has been well-received because it preserves the accuracy of plaintext computations perfectly, however, it still faces significant latency issues.
To address this, DELPHI~\cite{mishra2020delphi} splits the protocol into an offline phase for preprocessing and an online phase dependent on input values, significantly reducing online costs.
PRIMER~\cite{zheng2023primer} further enhanced it for Privacy-preserving inference of Transformer models (PiT), which focused on reducing the latency of HE but didn't address those of GC. Consequently, GC has emerged as the primary bottleneck, leaving its reduction as an unresolved challenge.

A primary strategy for effectively reducing latency in GC involves minimizing workload, particularly by reducing the number of AND gates within the circuit. Previous studies~\cite{testa2019reducing, testa2020logic, liu2022don} proposed methods to minimize the number of AND gates by transforming the circuit into a Directed Acyclic Graph (DAG), where each node represents a gate, and then analyzing this graph. However, they overlooked the fundamental approach of altering the structure of the circuit itself before transforming it into a DAG.

Another efficient method to reduce the latency of GC is utilizing a hardware accelerator. Although prior works~\cite{hussain2019fase,mo2023haac} have accelerated GC operations, the challenges remain while operating the nonlinear functions of transformers. FASE~\cite{hussain2019fase} assumes that all necessary data can be contained on-chip, making it impractical to compute with a vast amount of data. HAAC~\cite{mo2023haac} proposed a pipelined accelerator approach by suggesting the use of off-chip memory. Nevertheless, its scheduling scheme and on-chip memory policy are not optimal, resulting in significant memory stalls and pipeline stalls.

To summarize, previous studies have addressed protocols for PiT, reducing the workload of GC, and employing accelerators for GC, which is the primary bottleneck of the protocol. However, these approaches have only investigated parts of the issue and have not provided a holistic examination. Moreover, each of these schemes has failed to optimally reduce the overhead of GC when performing PiT. Hence, this paper introduces \sysname: a full-stack framework designed to accelerate PiT from the software level down to the hardware. \sysname makes the following contributions.
\begin{itemize}[topsep=5pt]
    \item A novel protocol that reduces GC latency by distributing workloads of the GC to alternative operations
    \item A GC-friendly circuit generation that significantly reduces the number of AND gates, thereby reducing both the GC latency of the online phase and the offline phase
    \item A new scheduling method that is composed of coarse-grained and fine-grained scheduling to minimize data dependency and increase data reuse
    \item A hardware accelerator with compiler speculation to improve data reusability and reduce redundant DRAM accesses
\end{itemize}
Overall, \sysname achieves a significant reduction in both offline and online latency by 2.2$\times$ and 12.2$\times$, respectively. Meanwhile, the APINT accelerator reduces its latency by 3.3$\times$ as well as lowers system energy consumption by 4.6$\times$ while operating PiT, compared to the state-of-the-art (SOTA) GC accelerator.
\section{Background and Motivation}

\begin{figure}[t]
    \vspace{-0.2in}
    \centering
    \includegraphics[width=1\linewidth]{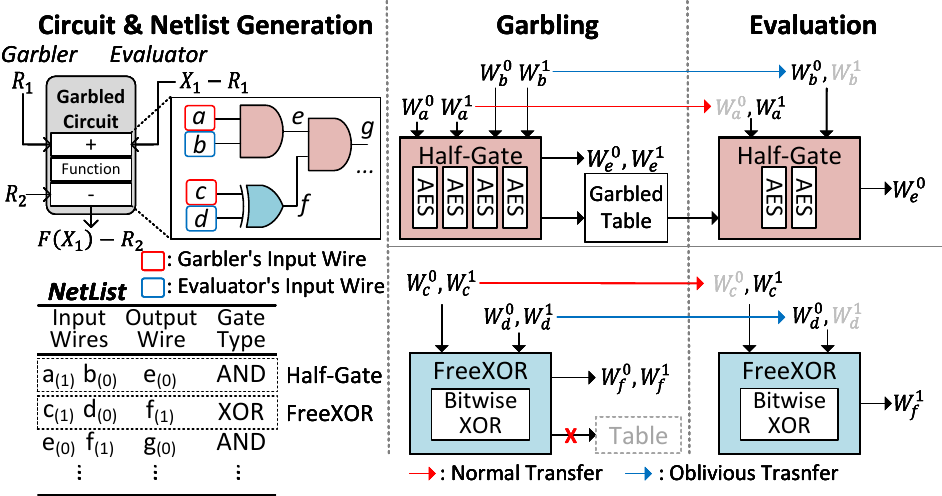}
    \caption{Garbled Circuit Protocol Overview}
    \vspace{-0.2in}
    \label{fig:garbled_circuit}
\end{figure}

\subsection{Protocols for Privacy-Preserving Inference}
\subsubsection{\textbf{Homomorphic Encryption}}
HE is a cryptographic method allowing operations on encrypted data (ciphertext) that produce the same results as if performed on unencrypted data (plaintext). It supports operations like addition and multiplication between plaintext and ciphertext ($Enc(X)+Y = Enc(X+Y)$, $Enc(X)\times Y = Enc(X\times Y)$), and among ciphertexts ($Enc(X)+Enc(Y) = Enc(X+Y)$, $Enc(X)\times Enc(Y) = Enc(X\times Y)$).



\subsubsection{\textbf{Garbled Circuits}}
GC involves two parties---the Garbler and the Evaluator---jointly compute without disclosing their inputs. The GC process involves four steps, as shown in Figure~\ref{fig:garbled_circuit}.\\
\textit{1) Circuit and Netlist Generation: } 
The function to be computed is represented as a circuit of 2-input logic gates, and it is converted to a netlist format, which contains information about each gate's inputs (input wires), output (output wire), and type (ex. AND).\\
\textit{2) GC Garbling: } 
For each wire \textit{i} in the netlist, the garbler randomly generates a 128b label ($W_i^0$) corresponding to 0. The label ($W_i^1$) corresponding to 1 is then produced by performing an XOR operation on ($W_i^0$) with a random 128b value R, which is fixed and public. After assigning the labels, the garbler constructs a truth table for each logic gate and encrypts it to create the garbled table. Finally, the garbler selects the corresponding label for its input wire ($a \in (W_a^0, W_a^1)$), which will be used in GC evaluation.\\
\textit{3) Garbler-Evaluator Communication:}
The garbler transmits the garbled tables and the selected label to the evaluator. Additionally, the garbler sends the label, corresponding to the evaluator's input wire ($b \in (W_b^0, W_b^1)$), to the evaluator without knowing the wire's value. This can be achieved through Oblivious Transfer (OT) protocol ~\cite{ishai2003extending}.\\
\textit{4) GC Evaluation:}
The evaluator calculates the output for each gate using the garbled tables and the labels provided by both the garbler and evaluator. The output becomes the new label of the input wire of the next gates, and the evaluations proceed sequentially for all gates in the netlist.

Recent enhancements in GC---Half-Gate~\cite{zahur2015two} for AND gates and FreeXOR~\cite{kolesnikov2008improved} for XOR gates---have reduced computational demands and memory footprints. To facilitate this, the netlist employs solely AND, XOR, and INV gates. It should be noted that INV gates, unlike AND and XOR gates, can be implemented at no cost by simply removing them and inverting the correspondence between the values and labels of the wires. In the garbling phase, contrary to the original GC that assigns labels of output wires randomly, the Half-Gate operation for AND gates produces an output wire's label and two garbled tables through four AES computations of the input wires' labels. Conversely, the FreeXOR operation for XOR gates generates an output wire's label by XOR computation of the input wires' labels without creating any garbled tables.
During the evaluation phase, a Half-Gate operation produces an output wire's label by garbled tables and two AES computations of input wires' labels, and a FreeXOR operation computes an output wire's label directly through an XOR operation of input wires' labels.

\begin{figure}[t]
    \vspace{-0.2in}
    \centering
    \includegraphics[width=1\linewidth]{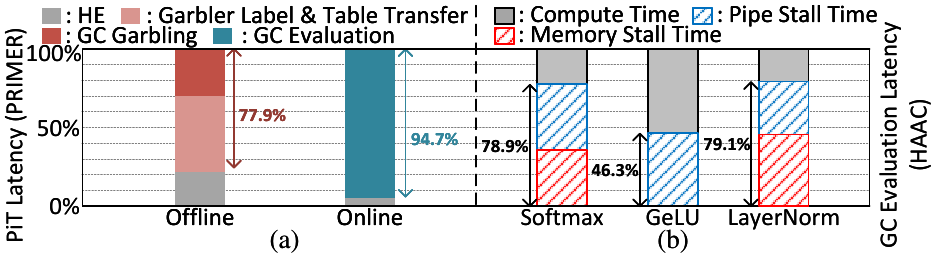}
    \caption{Latency Analysis of Prior Works (a) PRIMER Protocol and (b) HAAC}
    \vspace{-0.2in}
    \label{fig:APINT_motivation}
\end{figure}
    
\subsection{Motivation for \sysname}

Previous studies have several issues in effectively processing PiT by speeding up GC, the primary bottleneck.
First, in terms of protocol, PRIMER~\cite{zheng2023primer} has solely focused on optimizing HE and has not introduced any methods to reduce the latency of GC. As depicted in Figure~\ref{fig:APINT_motivation} (a), which presents the latency result for a single inference of the BERT-base model~\cite{devlin2018bert} with 128 tokens using the PRIMER protocol on a CPU, GC evaluation accounts for 94.7\% of the online latency. Moreover, GC garbling, along with the transfer of labels and garbled tables, contributes to 77.9\% of the offline latency. Therefore, there is a significant need for a new protocol that substantially reduces the latency of GC.

Second, regarding circuit generation, prior works~\cite{testa2019reducing, testa2020logic, liu2022don} haven't proposed the optimal way to reduce the number of AND gates. Given that Half-Gate operations are more complex than FreeXOR operations, reducing the number of AND gates can significantly decrease the GC workload. To achieve this, previous works have optimized the XOR-AND-Graph (XAG) in which the circuit is converted into a DAG. However, these approaches failed to realize a more optimal method by overlooking modifications to the fundamental implementation of the circuit.

Third, with respect to the accelerator, previous works~\cite{hussain2019fase, mo2023haac} mitigated the latency issue of GC, but none of them proposed practical solutions to operate the nonlinear functions of transformers. FASE~\cite{hussain2019fase}, which benchmarked with a small dataset, assumed that the on-chip itself could cover all the wires required during the operation.
However, this approach is infeasible since the nonlinear functions have many more wires than the on-chip can physically support. Although HAAC~\cite{mo2023haac} partially solved this problem with an additional scheme utilizing off-chip memory, it suffers from memory stalls and pipeline stalls as described in Figure~\ref{fig:APINT_motivation} (b) due to suboptimal scheduling, inefficient on-chip memory policies, and hardware structures not considering wire reusability.

\sysname aims to overcome these deficiencies by providing the full stack solution while fulfilling four key requirements. First, it must adhere to a new protocol that reduces the GC computations, the main bottleneck of PiT. Second, it should generate the circuit in a GC-friendly way that ensures a reduction of GC workloads to decrease latency and memory footprint. Third, it requires appropriate scheduling to reduce memory stalls and pipeline stalls. Fourth, it necessitates an accelerator equipped with the compiler, which resolves memory bottlenecks and redundant DRAM traffic.

\section{\sysname Framework}

\begin{figure}[t]
    \vspace{-0.2in}
    \centering
    \includegraphics[width=1\linewidth]{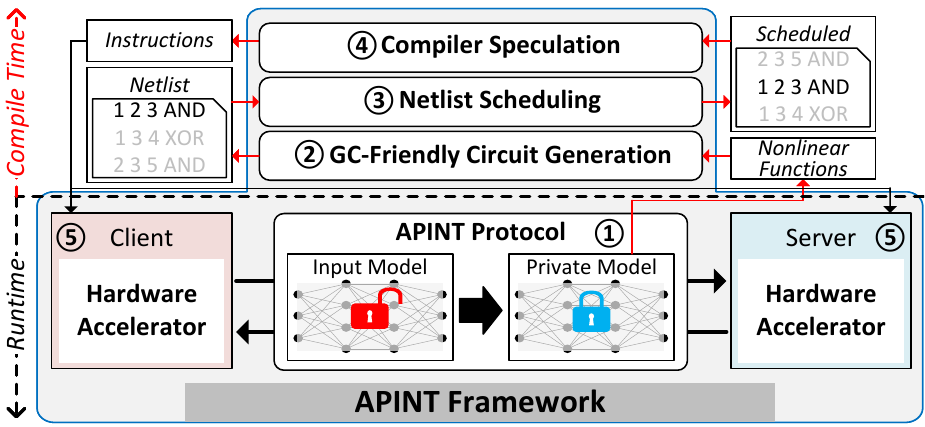}
    \caption{Overall \sysname Framework}
    \vspace{-0.2in}
    \label{fig:APINT_framework}
\end{figure}

In this section, we propose APINT, a full-stack framework designed to accelerate PiT by reducing the overhead of GC, the primary bottleneck of PiT. The overall workflow is illustrated in Figure~\ref{fig:APINT_framework}, distinguishing between compile-time and runtime processes.
In the initial compile time stage, \sysname begins by extracting nonlinear function operations in the process of computing the transformer model through the \sysname protocol. Next, through the GC-friendly circuit generation, the extracted function is implemented as a circuit consisting of a 2-input gate, and it is converted to netlist in Bristol format~\cite{tillich2016circuits}.
This step significantly alleviates the computational load on GC in subsequent stages. Following this, \sysname adopts a scheduling strategy that combines coarse-grained and fine-grained scheduling. The strategy enables full utilization of DRAM bandwidth and decrement in wire dependency. Furthermore, \sysname incorporates compiler speculation to generate instructions that capitalize on wire reusability and are executed on hardware accelerators at runtime. These accelerators, designed to further reduce memory stalls by eliminating redundant DRAM accesses, are deployed on both the server and client, allowing them to perform GC evaluation or GC garbling.

\subsection{\sysname Protocol}

\begin{figure}[t]
    \vspace{-0.2in}
    \centering
    \includegraphics[width=1\linewidth]{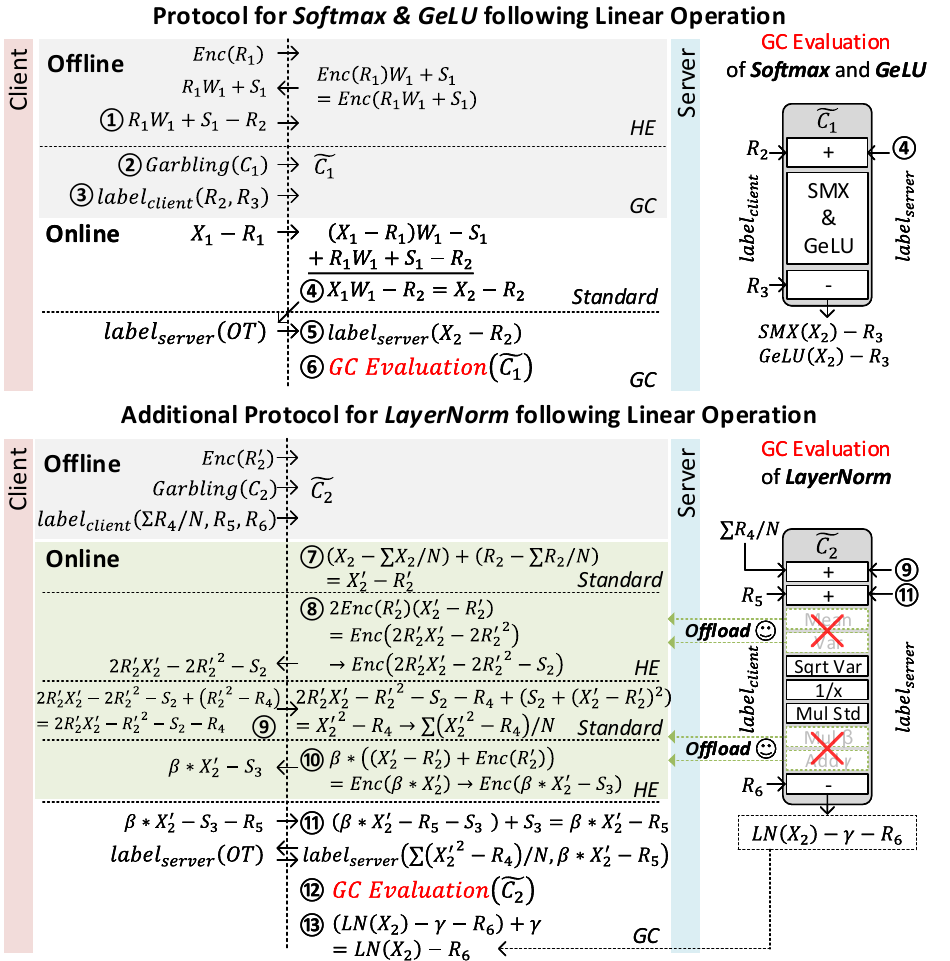}
    \caption{\sysname Protocol}
    \vspace{-0.2in}
    \label{fig:APINT_protocol}
\end{figure}

\sysname protocol is based on the PiT protocol in PRIMER, but it reduces the circuit in GC operation by offloading its partial calculations to HE and standard operations, thereby significantly decreasing the workload of GC. As illustrated in Figure~\ref{fig:APINT_protocol}, the basic concept of the protocol is the combination of HE for linear operations and GC for nonlinear operations. To maintain confidentiality, each party adds or subtracts a random matrix ($R_i$ of the client and $S_i$ of the server) before sending the data to each other.
At the offline phase, HE is utilized to compute linear function for the client's random matrix $R_{1}$, which has the same matrix size as the input matrix $X_1$ \circled{1}. Simultaneously, the client garbles the circuit $\Tilde{C_1}$, which integrates adding the secret shares from both parties, processing the nonlinear function, and subtracting a random matrix to ensure confidentiality \circled{2}. The client then transmits labels of $R_2, R_3$ to the server \circled{3}. During the online phase, the server calculates the linear function of $(X_{1}-R_{1})$ using standard matrix operations. This intermediate result is merged with data from process \circled{1} to complete the computation of the linear function, yielding $(X_{2}-R_{2})$ \circled{4}. After that, the labels of $(X_{2}-R_{2})$ are sent from the client via the OT protocol~\cite{ishai2003extending} \circled{5}, and the server proceeds the GC evaluation for the garbled $\Tilde{C_1}$ \circled{6}.

However, in contrast to other functions, the reduced circuit $\Tilde{C_2}$ is employed when operating LayerNorm. This circuit specifically excludes calculations of mean and variance, as well as operations involving the parameters $\beta$ and $\gamma$. The excluded calculations are offloaded and computed using standard operations and HE, thereby reducing the workload of GC.
During the offline phase, the client garbles the circuit $\Tilde{C_2}$, transmitting the labels for $\sum{R_4/N}, R_5, R_6$, and also sends $Enc(R_2')$.
During the online phase, the mean of $(X_{2}-R_{2})$ is initially calculated using standard operations. This mean is then subtracted from $(X_{2}-R_{2})$, resulting in $(X_{2}'-R_{2}')$ \circled{7}. Subsequently, to prepare for variance calculations, this result is multiplied by two times $Enc(R_2')$ \circled{8}. The results from equation \circled{7} and \circled{8} are then used to compute the variance \circled{9}. Third, multiplying with the parameter $\beta$ can be processed by utilizing HE with $(X_{2}'-R_{2}')$, $Enc(R_2')$, and $\beta$ \circled{10}, \circled{11}. Then, the labels of the data from the process \circled{9} and \circled{11}, which are obtained from the client via OT protocol, and the labels from the client are computed through GC evaluation \circled{12}. Finally, a straightforward addition of the parameter $\gamma$ is processed \circled{13}.

Although this protocol incurs additional overhead due to HE and communication of two parties, it brings a substantial reduction of GC latency, offsetting these increased costs perfectly. This reduction marks a significant improvement over the baseline protocol, which merely utilized GC for processing the nonlinear functions. Ultimately, the \sysname protocol achieves a significant reduction in the online latency of GC operations, reducing it by 47.3\% during the LayerNorm computation.

\subsection{GC-friendly Circuit Generation}

To further minimize the workload of GC, \sysname proposes GC-friendly circuit generation of the nonlinear functions. It involves implementing each function to a circuit with 2-input AND, XOR, and INV gates while preserving the accuracy of computations. The process unfolds in two main steps.

The initial step focuses on minimizing the total number of gates in the circuit. Since GC processes the gates of the circuit sequentially, reducing the gate count directly lowers the overall computational load. For instance, in the implementation of Softmax, the method from i-BERT~\cite{kim2021bert} is adopted, which scales inputs by \textit{ln2}, thereby reducing the range of values and the number of required gates for exponential operations. The exponential operations are performed through combinational logic, which performs as a Look-Up Table (LUT) interpolation. For the GeLU function, LUT interpolation is utilized after clipping the input values within a range (-4, 4)~\cite{gupta2023sigma}. In LayerNorm, the conventional approach is employed without any approximation, as it doesn't incur any accuracy drop.

\begin{figure}[t]
    \vspace{-0.2in}
    \centering
    \includegraphics[width=1\linewidth]{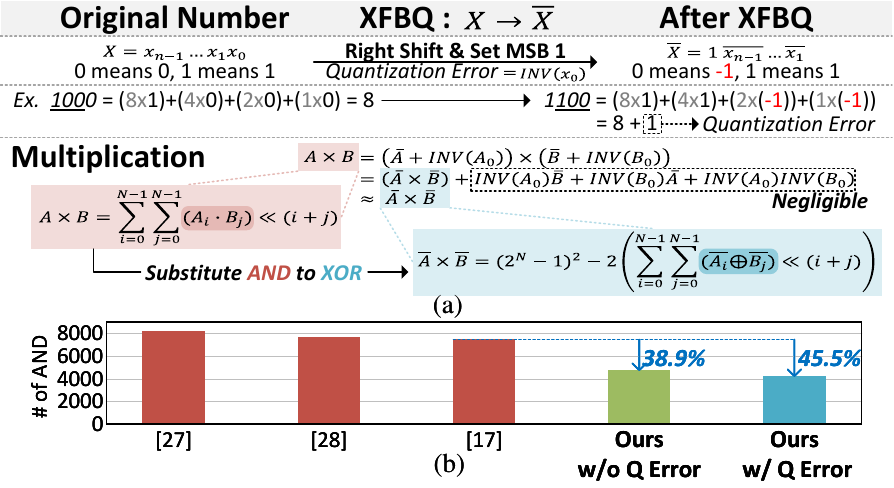}
    \caption{(a) Reduction of ANDs via XBFQ Multiplication (b) Comparison of ANDs for 64b Multiplication}
    \vspace{-0.2in}
    \label{fig:circuit_generation}
\end{figure}

The second step aims to decrease the number of AND gates further. \sysname proposes a method employing XOR-Friendly Binary Quantization (XFBQ)~\cite{jian2020fast} to implement multiplication with fewer AND gates compared to the conventional method. This approach is motivated by the observation that the multiplication process accounts for a significant portion of each nonlinear function's implementation. 
Figure~\ref{fig:circuit_generation} (a) summarizes XFBQ and its multiplication process. XFBQ modifies the binary representation, wherein 1 represents 1 and 0 represents -1, exploiting the correspondence between the result patterns of XOR operations and the product of 1 and -1. The way to XFBQ is straightforward: it involves a right shift and changing the MSB to 1, introducing only a minimal quantization error (Q error) as small as the INV of Least Significant Bit (LSB) of the original number. For instance, \textit{1000}=8 turns into \textit{1100}=8+4-2-1=9 after XFBQ with Q error as \textit{INV(0)=1}.  When expressing conventional multiplication ($A\times B$) as XFBQ multiplication ($\hat{A} \times \hat{B}$) along with additional terms due to Q errors, the AND operations of the conventional method are all replaced by XOR, thereby a high reduction of AND gates occurs. Moreover, given that the Q error is INV of LSB value, its negligible impact on multiplication results and PiT warrants disregarding the additional terms, leading to a further decrease in AND gates.

Figure~\ref{fig:circuit_generation} (b) shows the effects of the multiplication using XFBQ while operating 64b multiplication. It reduces the number of AND gates 38.9-45.5\% compared to prior work~\cite{liu2022don}, depending on the inclusion of Q error adjustments. In addition, GC-friendly circuit generation employed methods from the work~\cite{testa2020logic} for operations other than multiplication, which has been proven to perform as an open-source. Finally, it reduces the workload of GC for nonlinear functions by an average of 42.5\%.

\begin{figure*}[t]
    \vspace{-0.2in}
    \centering
    \includegraphics[width=1\linewidth]{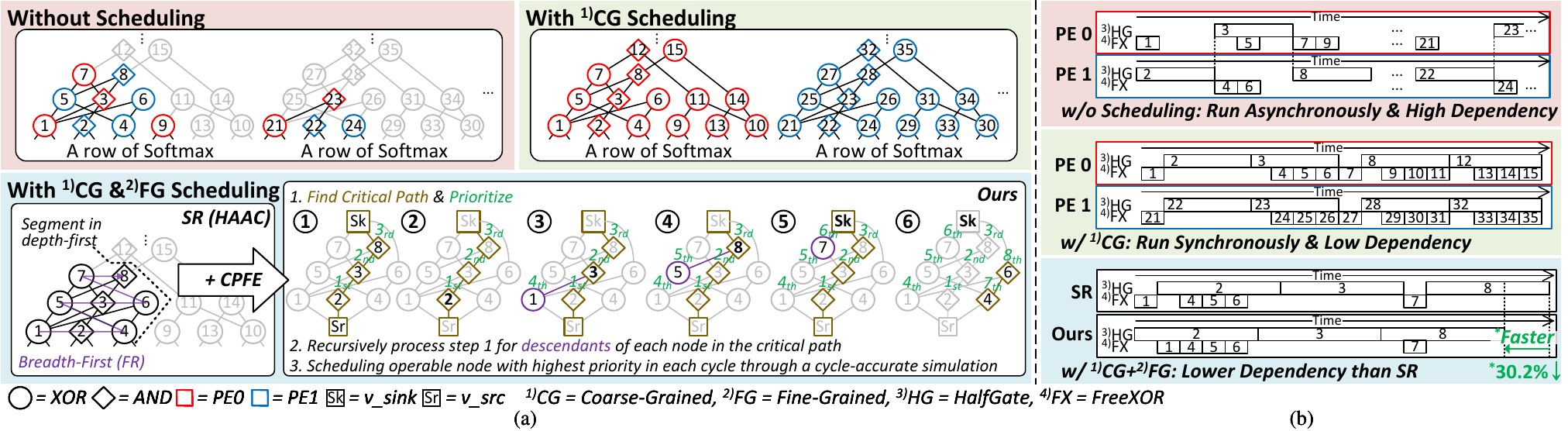}
    \caption{(a) Methods and (b) Effects of Coarse-Grained and Fine-Grained Scheduling}
    \vspace{-0.2in}
    \label{fig:scheduling}
\end{figure*}
\subsection{Netlist Scheduling}
Despite the reductions in GC overhead facilitated by the \sysname protocol and GC-friendly circuit generation, GC still accounts for a notable portion of the latency. This implies that reducing GC overhead necessitates the integration of hardware accelerators beyond software solutions.
However, since a GC accelerator takes a netlist, converted from the circuit, as input and processes gates in the netlist sequentially, efficient netlist scheduling is crucial for hardware acceleration. Therefore, we introduce coarse-grained and fine-grained scheduling, maximizing DRAM bandwidth utilization and minimizing computational dependencies to accelerate the GC operation of nonlinear functions. 


\subsubsection{\textbf{Coarse-Grained Scheduling}}

Due to the complex implementation of the circuit of the nonlinear function, the netlist exhibits highly irregular patterns in input and output wires, leading to irregular DRAM accesses that hinder optimal DRAM bandwidth utilization.
To tackle this issue, \sysname adopts coarse-grained scheduling, which maps each independent operation onto each core, allowing cores to function independently yet synchronously. This approach leverages the characteristic of nonlinear functions composed of independent unit operations, such as rows in Softmax, to be computed separately. 
Assuming there are two Processing Engines (PEs) and the need to compute two Softmax rows, the red box of Figure~\ref{fig:scheduling} (a) illustrates a DAG of two rows ordered in a depth-first manner without any scheduling, where node number corresponds to the order of the gates in the netlist. In this case, two PEs concurrently process the netlists of two rows in a dependent manner. In contrast, as illustrated in the green box, each PE exclusively handles an independent row with coarse-grained scheduling. Hence, contrary to the red box in Figure~\ref{fig:scheduling} (b), the green box demonstrates that coarse-grained scheduling allows all PEs to operate synchronously, ensuring they request DRAM data simultaneously. Therefore, while the intra-core DRAM access pattern is irregular, the inter-core DRAM access pattern becomes the same. As a result, by enabling cores to share the DRAM data bus, coarse-grained scheduling ensures maximal utilization of DRAM bandwidth.

Moreover, coarse-grained scheduling offers the additional benefit of resolving wire dependencies. Unlike the scenario without coarse-grained scheduling, the scheduling enables each PE to independently operate on a distinct row without dependencies. Thus, coarse-grained scheduling not only maximizes the utilization of DRAM bandwidth but also reduces the pipeline stalls.
        
\subsubsection{\textbf{Fine-Grained Scheduling}}
In addition to the coarse-grained scheduling, \sysname applies fine-grained scheduling that further diminishes GC latency compared to Full Reorder (FR) and Segment Reorder (SR), the scheduling method by the SOTA GC accelerator, HAAC~\cite{mo2023haac}. The FR transforms the netlist into a DAG and establishes processing order by traversing the graph in a breadth-first manner, reducing wire dependencies. However, due to limited on-chip memory, it can cause off-chip traffic by spilling wires over to DRAM, especially in applications where the DAG has a wide breadth, such as the nonlinear function of transformers. To tackle this problem, HAAC proposed the SR that segments the netlist, ordered in a depth-first manner, to enhance wire reuse and then applies FR within each segment to reduce wire dependencies. 
Despite these advancements, we identified opportunities for further latency reduction since FR does not optimally eliminate wire dependency within segments. Therefore, \sysname introduces a fine-grained scheduling strategy that combines segmentation and Critical-Path-First-Execution (CPFE)~\cite{zhao2020dag, zhao2022dag} instead of FR, achieving enhanced performance by effectively minimizing wire dependencies.

The fine-grained scheduling begins by segmenting the netlist, with each segment half the size of on-chip memory. A DAG is then constructed for each segment, with assigning weights reflecting the cycle latency of each gate. Nodes without children are linked to \textit{v\_{src}}, and those without parents are connected to \textit{v\_{sink}}. After establishing the DAG, it finds a critical path from \textit{v\_{src}} to \textit{v\_{sink}} and prioritizes nodes along this path, starting from the lowest depth. Subsequently, for each node on the path, a sub-DAG is formed comprising unprioritized descendants, and the process of identifying the critical path and assigning priorities repeats recursively.

The blue box in Figure~\ref{fig:scheduling} (a) shows how the fine-grained scheduling works. First, in step \circled{1}, it finds a critical path and prioritizes from the lowest depth. Then, from step \circled{2}-\circled{6}, it creates sub-DAG with unprioritized descendants for each node of the path and operates recursively. 
For example, in step \circled{6}, nodes \textit{4} and \textit{6} compose the sub-DAG, and then the process of identifying the critical path and prioritizing is recursively executed at the sub-DAG. After assigning priorities to all nodes, the scheduling order is determined by the cycle-accurate simulation. The simulation selects the operable node with the highest priority in each cycle. The "operable" refers to the condition where both input wires of a DAG node have been produced. As a result, step \circled{6} is reordered as $2\rightarrow1\rightarrow4\rightarrow5\rightarrow6\rightarrow3\rightarrow8\rightarrow7$. 
Hence, as depicted in Figure~\ref{fig:scheduling}, fine-grained scheduling significantly reduces pipeline stalls by wire dependencies within each segment, enhancing the computation speed of nonlinear functions by an average of 30.2\% compared to the SR of HAAC.

\begin{figure*}[t]
    \vspace{-0.2in}
    \centering
    \includegraphics[width=1\linewidth]{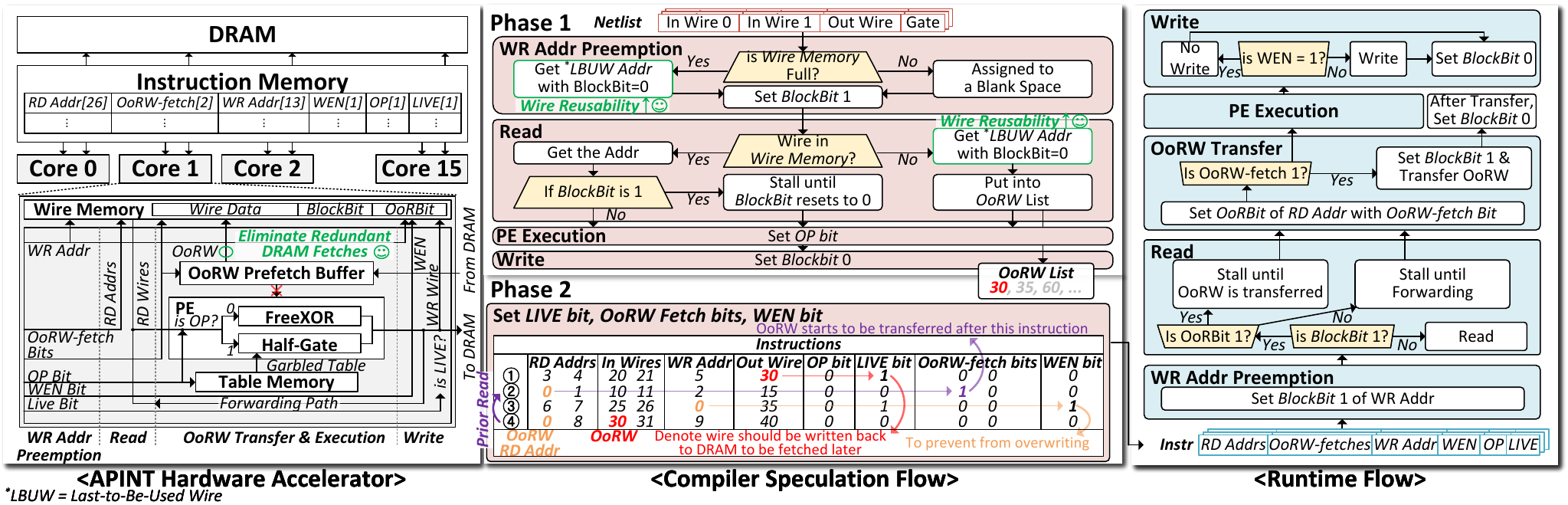}
    \caption{APINT hardware, Compiler Speculation Flow, and Runtime Flow Descriptions}
    \vspace{-0.2in}
    \label{fig:compiler_and_hardware}
\end{figure*}
\subsection{Accelerator with Compiler Speculation}
HAAC introduced an accelerator showcasing superior performance compared to CPUs and GPUs by utilizing pipelined multi-core designs operating concurrently and leveraging off-chip memory when on-chip memory resources are insufficient. However, it encounters a significant memory bottleneck when processing nonlinear functions of transformers due to inadequate on-chip memory policy and hardware structure.
HAAC's approach of sequentially writing output wires in on-chip memory doesn't consider the wire reusability. Moreover, the hardware structure that involves directly fetching wires from DRAM to a PE via a queue structure limits the wire's usage to a single time, thereby restricting its potential for reuse. To counter these issues, \sysname suggests the accelerator alongside compiler speculation techniques, aiming to reduce unnecessary DRAM accesses and improve wire reuse.

\subsubsection{\textbf{APINT Accelerator}}
Figure~\ref{fig:compiler_and_hardware} illustrates the architecture of \sysname accelerator, which features 16 independent cores operating synchronously under coarse-grained scheduling. This eliminates the need for inter-core communication, allowing for a shared unified Instruction Memory (16KB). Each core includes a Wire Memory (128KB), a Table Memory (2KB), an Out-of-Range-Wire (OoRW) Prefetch Buffer (1KB), and a PE, all of which are pipelined.
Wire Memory stores the wire's label (value of the wire) and special flag bits, which are a block bit and an Out-of-Range (OoR) bit, for each address. The block bit prevents other wires from accessing the address, while the OoR bit indicates that an OoRW, a wire that is fetched from DRAM, is being fetched to that address. Also, the Table Memory stores garble tables required for Half-Gate operations, and the OoRW Prefetch Buffer temporarily stores OoRWs fetched from DRAM. They are then transferred to the Wire Memory, which allows multiple reuses within the Wire Memory in contrast to HAAC, where OoRWs are used only once per fetch.

The execution of the accelerator is structured into four stages. First, upon receiving an instruction, the Write Address Preemption stage the write address in the Wire Memory, activating the block bit. Next, the Read stage reads two input wires from the memory or forwarding path over three cycles. Third, the input wires are processed in the Half-gate unit (taking 18 cycles for evaluation and 21 for garbling) or FreeXOR unit (taking one cycle) in PE, and OoRWs are transferred from the Prefetch Buffer to Wire Memory if required. Finally, the output wire generated in the PE is written back to Wire Memory over two cycles and, if needed, also to DRAM.


\subsubsection{\textbf{Compiler Speculation Flow}}
Before running the accelerator, compiler speculation is initially processed with a netlist as input. Its purpose is to generate instructions for the accelerator, which implements a memory policy that enhances wire reuse. It proceeds through the following two phases, as depicted in Figure~\ref{fig:compiler_and_hardware}. During the first phase, it assigns read and write addresses in Wire Memory and an OP bit for each gate in the netlist through a cycle-accurate simulation. After filling Wire Memory as much as possible with operable input wires, the speculation begins with the Write Address Preemption stage, allocating a write address either to a blank space or to the Last-to-Be-Used Wire (LBUW), if Wire Memory is full. The LBUW is the wire that will be used last among wires within the memory. This demonstrates that APINT employs a memory policy considering the reusability of wires. After assigning the write address, the block bit is activated for the preemption.


Next, the Read stage assigns the read address based on whether an input wire is present in Wire Memory. If present, the read address corresponds to its location. If not, indicating it is likely to become an OoRW at runtime, the address of the LBUW with inactive block bit is assigned, which also contributes to the memory policy that considers the reusability of wires. The input wire is then replaced with the LBUW and added to the OoRW list. Subsequently, the PE execution stage begins by setting the OP bit based on the gate type. This is followed by the Write stage, which writes the output wire at the assigned address and resets the block bit. This cycle-accurate simulation is repeated until every wire and gate in the netlist has been allocated the instructions with addresses and OP bits.

After completing the first phase, the second phase involves assigning the Live bit, two OoRW-fetch bits, and the Write Enable Not (WEN) bit, which are determined by analyzing the interrelationships among instructions. The Live bit is assigned to instructions that output an OoRW, designating that the wire should be written to DRAM for later use. For example, in Figure~\ref{fig:compiler_and_hardware}, instruction \circled{1} outputs OoRW \textit{30} and is marked with a Live bit of 1.
Each OoRW-fetch bit is assigned to ensure the timely transfer of an OoRW from the Prefetch Buffer to Wire Memory, based on instruction sequence and read dependencies.
For instance, instruction \circled{2}, which reads address \textit{0} immediately before instruction \circled{4} reads OoRW \textit{30} from the same address, is assigned an OoRW-fetch bit to ensure that OoRW \textit{30} is transferred right after instruction \circled{2} reads the address \textit{0} to prevent stalls due to non-arrival.
The WEN bit is assigned to prevent premature overwriting in Wire Memory.
For example, if OoRW \textit{30} is transferred to memory before instruction \circled{3} writes to address \textit{0}, it could be overwritten before it is read by instruction \circled{4}. Therefore, a WEN bit is assigned to instruction \circled{3} to prevent it from overwriting OoRW \textit{30}, and wire \textit{35} is written only to DRAM as dictated by the Live bit.


\subsubsection{\textbf{Runtime Flow}}

During the speculation process, the DRAM addresses for instructions, garbled tables, and OoRWs are predetermined, removing the need to handle the addresses during runtime. While fetching the data from DRAM to each corresponding memory, the runtime process is executed in the following four stages, as depicted in Figure~\ref{fig:compiler_and_hardware}.
After an instruction is decoded, the Write Address Preemption stage activates the block bit at the write address, and the Read stage operates based on the statuses of the block and OoR bits. Depending on these bits, the accelerator either performs a normal read or stall until the necessary wire is transferred from the Prefetch Buffer or the forwarding path. The OoRW Transfer and PE Execution stage then commences. The OoR bit is assigned with the OoRW-fetch bit, indicating whether an OoRW transfer is started. If the OoRW-fetch bit is 1, the address is preempted by activating the block bit, and an OoRW begins to be transferred to the address. After the completion of the transfer, the block bit is deactivated. Concurrently, the PE processes Half-gate or FreeXOR operation based on the OP bit. After the output wire is generated in the PE, the Write stage begins, writing the wire to Wire Memory and DRAM depending on the WEN and Live bits. After these stages are executed across all instructions, the runtime process is completed. Overall, through APINT accelerator and compiler speculation, it achieves a reduction in memory stall times by 86.1\% to 99.4\% compared to HAAC when operating nonlinear functions.

\section{Evaluation}

\subsection{Evaluation Setup}

\begin{figure}[t]
\vspace{-0.12in}
\centering
\includegraphics[width=1\linewidth]{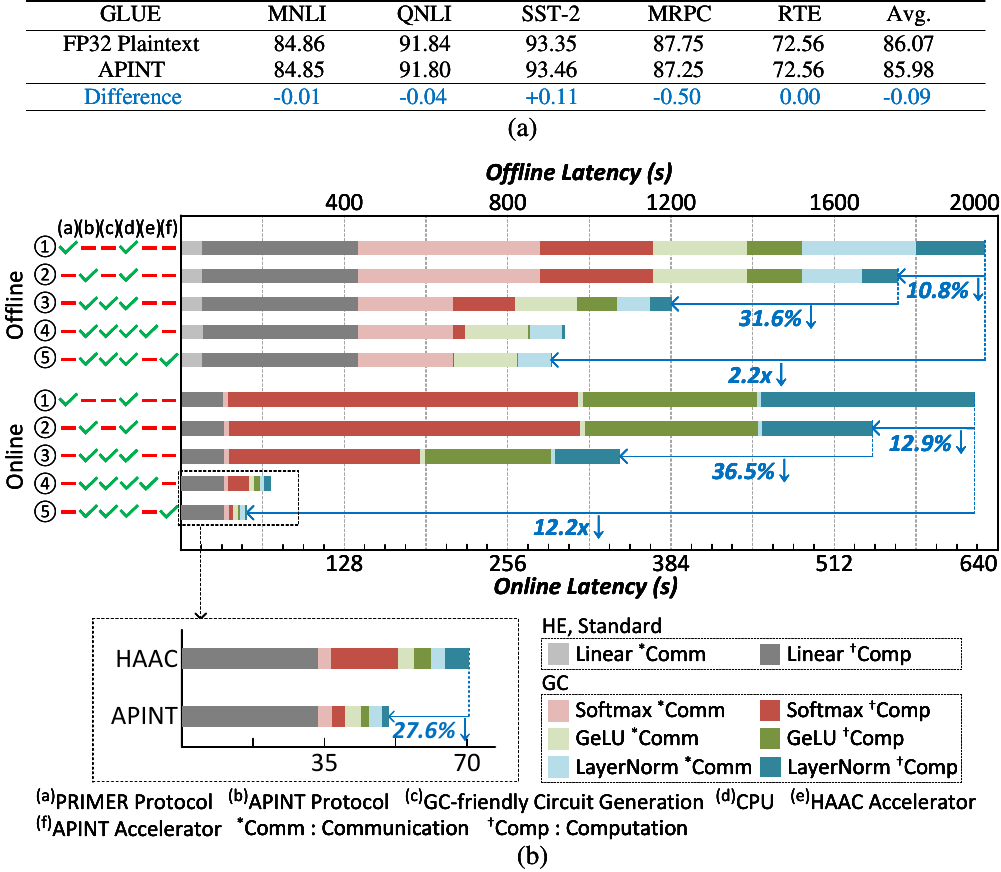}
\caption{A Breakdown of (a) Accuracy and (b) Latency of APINT in BERT-base Model Inference
with 128 Tokens}
\vspace{-0.2in}
\label{fig:protocol_latency}
\end{figure}

To evaluate the APINT protocol, we configured the two instances, one for a client and the other for a server, each with Intel Xeon Platinum 8452Y CPU~\cite{intelIntelXeon} at 2.0GHz, 32 threads, and 1TB of DRAM. They communicated via a network with 0.165ms latency and 9.6Gbps bandwidth, akin to LAN environments as prior study~\cite{boemer2020mp2ml}. CPU energy was tracked by a commercial tool~\cite{intelRunningAverage}. GC and HE(BFV~\cite{brakerski2014leveled}) were implemented using EMP-tool~\cite{emp-toolkit} and SEAL~\cite{sealcrypto} libraries, respectively. The security level of GC is verified by AES-128. The benchmark utilized BERT-base~\cite{devlin2018bert} with 128 tokens and five datasets from GLUE~\cite{wang2018glue}. The bit precision was set at 37 bits for Softmax and LayerNorm, and 21 bits for GeLU, ensuring accuracy as recent study~\cite{pang2023bolt}.
The constituent circuits of the functions were implemented via Verilog and then converted into a netlist using the Synopsys Design Compiler~\cite{synopsysDesignCompiler}. These netlists were subsequently transformed into Bristol format and merged using EMP-tool.

A cycle-accurate simulator was developed to model the GC accelerator of \sysname and HAAC, with running memories at 2GHz and computational units at 1GHz. 
To accurately measure memory stalls caused by external DRAM accesses, we used Ramulator~\cite{kim2015ramulator}, the well-known DRAM simulator, after configuring it with HBM2 specifications~\cite{o2017fine}. Also, for HBM2's energy consumption measurement, we utilized values from prior research~\cite{o2017fine}.
For the computational units, we measured the area by synthesizing using 28nm technology via Synopsys Design Compiler. For their energy consumption measurement and timing verification, we utilized the PrimeTime~\cite{synopsysGoldStandard}.
In addition, most of the memories are constructed using SRAMs, except for some registers storing special bits. Specifically, the SRAM's specification was taken from the TSMC N28HPC+ Memory Compiler~\cite{usermanual, mo2023haac}, aligning with HAAC's conditions. 
For comparative analysis with HAAC, the total area was scaled to 16nm, applying a reduction factor of 1.9x, following HAAC’s methodology~\cite{statnano16FFFinFET, mo2023haac, statnanoTSMC20nm}.

\begin{figure} [t]
\vspace{-0.12in}
\centering
\includegraphics[width=1\linewidth]{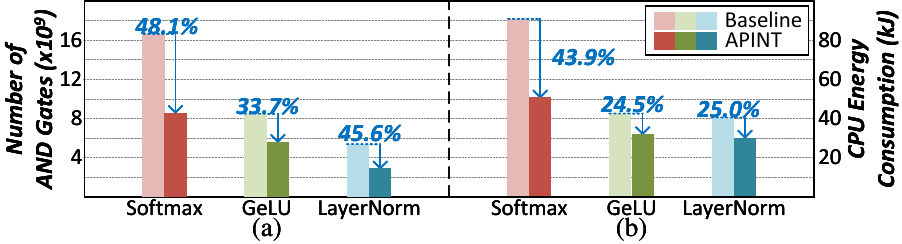}
\caption{The Reduction of (a) ANDs and (b) CPU Energy Consumption by GC-friendly Circuit Generation}
\vspace{-0.2in}
\label{fig:netgen_result}
\end{figure}

\subsection{APINT Protocol}

\begin{figure*}[t]
    \vspace{-0.2in}
    \centering
    \includegraphics[width=1\linewidth]{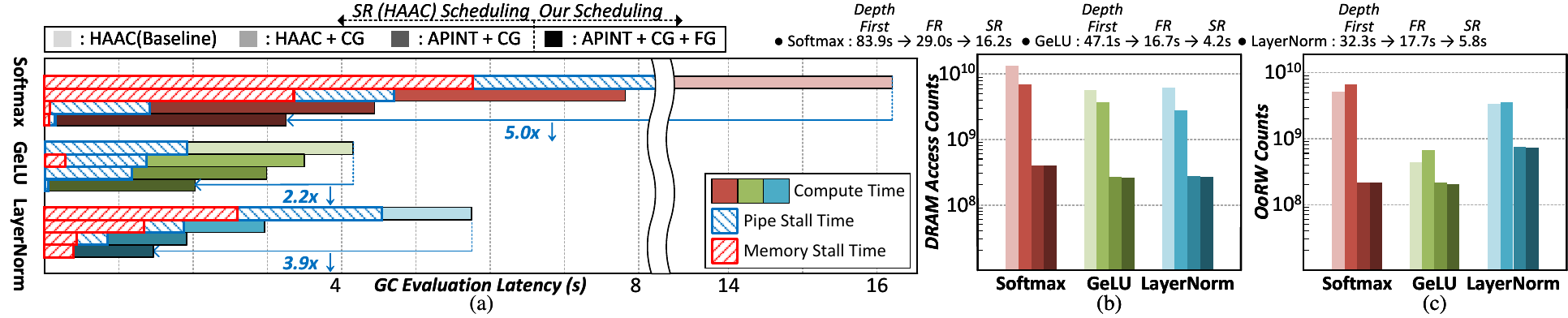}
    \caption{Breakdown of (a) Latency, (b) OoRW Counts, and (c) DRAM Access Counts for GC Evaluation of Nonlinear functions in BERT-base Model Inference
with 128 Tokens across Scheduling, Speculation, and GC Accelerators}
    \label{fig:total_eval}
\end{figure*}

\subsubsection{\textbf{Accuracy}}
Figure~\ref{fig:protocol_latency} (a) shows the accuracy of a BERT-base model using APINT protocol on the five datasets from GLUE, compared with plaintext FP32. The results demonstrate that the \sysname protocol is feasible for performing PiT while maintaining accuracy.

\subsubsection{\textbf{Latency}}
\circled{1} and \circled{2} in Figure~\ref{fig:protocol_latency} (b) displays latency reductions for a single inference in an end-to-end BERT-base model processing 128 input tokens, comparing the \sysname protocol with the PRIMER's protocol, both executed on a CPU. During the offline phase, computation latency arises from GC garbling, and communication latency is due to the transfer of the circuit and client labels. In the online phase, computation latency results from GC evaluation, and communication latency is caused by OT. The \sysname protocol reduces the circuit for LayerNorm operation, leading to a 10.8\% reduction in circuit transmission and GC garbling during the offline phase, and a 12.9\% reduction in the GC evaluation in the online phase.

\subsection{GC-friendly Circuit Generation}

\subsubsection{\textbf{Latency}}
As depicted in Figure~\ref{fig:netgen_result} (a), APINT's GC-friendly circuit generation employs XBFQ multiplication to reduce the number of AND gates, achieving reductions of 48.1\% for Softmax, 33.7\% for GeLU, and 45.6\% for LayerNorm for a single inference, compared to the baseline method by Testa~\cite{testa2020logic}. This reduction directly corresponds to a decrease in Half-Gate operations, subsequently reducing both the GC garbling and evaluation latency. In addition, it decreases the number of garbled tables, thereby reducing the communication latency of transferring them during the offline phase. Consequently, as shown in \circled{3} in Figure~\ref{fig:protocol_latency} (b), an additional application of the GC-friendly circuit generation to the APINT protocol results in a 31.6\% and 36.5\% reduction in latency during the offline and online phases, respectively.

\subsubsection{\textbf{CPU Energy Consumption}}
The reduction in Half-Gate operations also impacts the CPU's energy consumption. As depicted in Figure~\ref{fig:netgen_result} (b), the circuit generation reduces the energy consumption of a CPU for a single inference by 43.9\% for Softmax, 24.5\% for GeLU, and 25.0\% for LayerNorm. To sum up, APINT's GC-friendly circuit generation not only reduces latency during both online and offline phases but also lowers energy consumption.

\subsection{Netlist Scheduling, Compiler Speculation, and APINT Accelerator}

\subsubsection{\textbf{Latency}}
Figure~\ref{fig:total_eval} depicts the reduction of latency, OoRWs, and DRAM accesses achieved by the netlist scheduling, compiler speculation, and APINT accelerator while operating nonlinear functions in a single inference, compared to the HAAC. However, when using a netlist ordered by EMP-tool's depth-first scheduling and a netlist scheduled by FR scheduling of HAAC, significant latency occurs due to high dependency and substantial memory stalls, respectively. Therefore, their results are only briefly mentioned at the top of the graph, and the \sysname accelerator was evaluated against the HAAC accelerator using SR scheduling.

\textit{\textbf{Coarse-grained Scheduling}} Coarse-grained scheduling significantly reduces pipeline stalls across all nonlinear functions by resolving wire dependencies. It also decreases memory stall time for Softmax and LayerNorm, yet it marginally increases it for GeLU. This variation is due to a trade-off of the scheduling approach. As illustrated in Figure~\ref{fig:total_eval} (b), the schedule enhances DRAM bandwidth utilization, resulting in lower DRAM access counts. Conversely, as depicted in Figure~\ref{fig:total_eval} (c), it causes an increase in OoRWs due to the segmented allocation of on-chip memory across different cores. This effect is particularly pronounced for GeLU, which has a higher number of operable independent wires compared to other functions. Thus, memory stalls are slightly increased for GeLU, unlike other functions. Nonetheless, the reduction in pipeline stall times dominates in a decrease in total latency for all functions.

\textit{\textbf{APINT Accelerator \& Compiler Speculation}} APINT accelerator with compiler speculation notably diminishes memory stalls across all functions compared to HAAC's accelerator with course-grained scheduling, as illustrated in Figure~\ref{fig:total_eval} (a). This result stems from a substantial reduction in the number of OoRWs due to compiler speculation and a significant decrease in unnecessary DRAM accesses enabled by the OoRW fetching approach of the APINT accelerator, as depicted in Figure~\ref{fig:total_eval} (b) and (c). Therefore, unlike HAAC, \sysname accelerator with compiler speculation has resolved most of the memory bottleneck.

\textit{\textbf{Fine-grained Scheduling}} Applying fine-grained scheduling significantly reduces pipeline stalls compared to employing HAAC's SR scheduling, demonstrating the superiority of APINT's scheduling in resolving wire dependencies, as described in Figure~\ref{fig:total_eval} (a). This conclusively shows that \sysname not only resolves memory bottlenecks but also optimally settles the wire dependencies.

In summary, the APINT accelerator resolves memory bottlenecks and further enhances the compute performance, achieving an average latency reduction of 3.3$\times$ across all functions, with a specific reduction of 5.0$\times$ for Softmax, 2.2$\times$ for GeLU, and 3.9$\times$ for LayerNorm during GC evaluation compared to HAAC. Given that the only difference between GC garbling and GC evaluation is an additional three cycles required to compute the Half-Gate, similar latency reduction ratios are observed in both processes. This improvement of the APINT accelerator results in a 27.6\% reduction in online latency, while HE is processed by CPU, as depicted in \circled{4} and \circled{5} of Figure~\ref{fig:protocol_latency}. Overall, APINT's full-stack solutions result in a substantial reduction in latency, with a 2.2$\times$ decrease during the offline phase and a 12.2$\times$ reduction during the online phase.

\begin{figure} [t]
\vspace{-0.12in}
\centering
\includegraphics[width=1\linewidth]{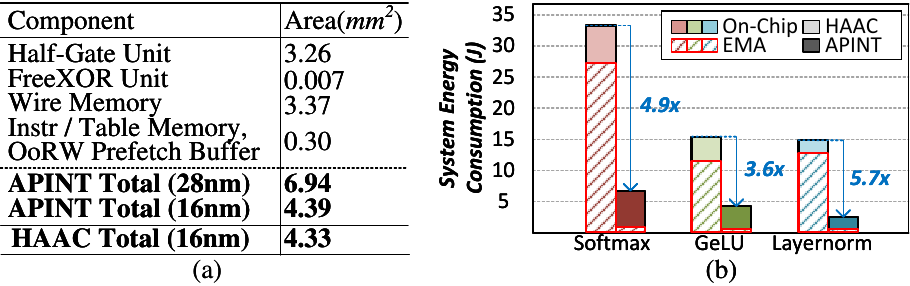}
\caption{A Breakdown of (a) Area of APINT Accelerator and (b) System Energy Consumption}
\vspace{-0.2in}
\label{fig:power_result}
\end{figure}

\subsubsection{\textbf{System Energy Consumption}} The reduction in DRAM accesses not only alleviates memory stalls but also significantly lowers system energy consumption. Unlike HAAC, which primarily focuses on on-chip energy consumption, we highlight the importance of managing consumption due to external memory accesses (EMA). As depicted in Figure~\ref{fig:power_result}, although the APINT accelerator has nearly the same area and on-chip energy consumption as HAAC, it achieves a significant reduction in EMAs, leading to substantial system energy savings of 4.9$\times$ for Softmax, 3.6$\times$ for GeLU, and 5.7$\times$ for LayerNorm, averaging a 4.6$\times$ decrease overall.
\section{Conclusion}
In conclusion, this paper introduces APINT, the pioneering full-stack framework to accelerate PiT. APINT integrates a novel PiT protocol and GC-friendly circuit generation, effectively reducing the latency of GC. Additionally, APINT includes a netlist scheduling and an accelerator equipped with compiler speculation, resolving both the memory bottleneck and wire dependency while operating nonlinear functions of the transformer. As a result, APINT achieves a substantial reduction in latency, performing 12.2$\times$ and 2.2$\times$ faster online and offline, respectively, compared to the baseline method on CPU. 
Meanwhile, the APINT accelerator reduces its latency by an average of 3.3$\times$ and system energy consumption by 4.6$\times$ compared to the SOTA GC accelerator.

\bibliographystyle{ACM-Reference-Format}
\bibliography{ref}

\end{document}